\documentclass[runningheads]{llncs}
\usepackage{graphicx}
\usepackage{float}
\usepackage{pythontex}
\usepackage{pgfplots}
\usepackage{dsfont}
\usepackage{bbm}
\usepackage{amsmath}
\usepackage{amssymb}

\pgfplotsset{compat=1.14}

\definecolor{mygray}{gray}{0.8}
\begin{document}

\title{Weighted Adaptive Coding}

\author{Aharon Fruchtman \inst{1}
\and
Yoav Gross \inst{1}
\and\\
Shmuel T.\ Klein\inst{2}
\orcidID{0000-0002-9478-3303} 
\and 
Dana Shapira \inst{1} \orcidID{0000-0002-2320-9064}}

\authorrunning{A.\ Fruchtman et al.}

\institute{Computer Science Department, Ariel University,  Israel 
\email{\{aralef,yodgimmel\}@gmail.com, shapird@g.ariel.ac.il}\\
\and
Computer Science Department, Bar Ilan University,  Israel\\
\email{tomi@cs.biu.ac.il}}

\maketitle             

\begin{abstract}
Huffman coding is known to be optimal, yet its dynamic version may be even more efficient in practice.
A new variant of Huffman encoding has been proposed recently, that provably always performs better than
static Huffman coding by at least $m-1$ bits, where $m$ denotes the size of the alphabet, and has a better worst case than the standard dynamic Huffman coding. This paper introduces a new generic coding method, extending the known static and dynamic variants and including them as special cases. 
In fact, the generalization is applicable to all statistical methods, including arithmetic coding.
This leads then to the formalization of a new adaptive coding method, which is provably always at least as good as the best dynamic variant known to date.
Moreover, we present empirical results that show  improvements
over static and dynamic Huffman and arithmetic coding
achieved by the proposed method,
even when the encoded file includes the model description.

\keywords{Data Compression \and Static and Dynamic Coding \and Huffman Coding \and Arithmetic Coding}

\end{abstract}




\section{Introduction}

Huffman coding \cite{Huf} is one of the seminal techniques in data compression and is applied on a set of elements $\Sigma$ into which a given input file $T$ can be partitioned. We shall refer to $\Sigma$ as an {\it alphabet\/} and to its elements as {\it characters\/}, but these terms should be understood in a broader sense, and the characters may consist of strings or words, as long as there is a well defined way to break $T$ into a sequence of elements of $\Sigma$.

Huffman coding is known to be optimal in case the alphabet is known in advance, the set of codewords is fixed and each codeword consists of an integral number of bits. If one of these conditions is violated, optimality is not guaranteed.
A recent comprehensive survey on Huffman coding is given in \cite{moffat}.


In the {\it dynamic\/} variant of Huffman coding, also known as {\it adaptive\/}, the encoder and decoder maintain identical copies of the model; at each position, the model consists of the frequencies of the elements processed so far.
After each processed element $\sigma$, the model is updated by incrementing the frequency of $\sigma$ by 1, while the other frequencies remain the same. 
Specifically, Faller \cite{Faller} and Gallager \cite{Gallager} propose a one-pass solution for dynamic Huffman coding. Knuth extends Gallager's work and also suggests that the frequencies may be decreased as well as increased \cite{Knuth}, which enables the usage of a sliding window rather than relying on the full history. These independent adaptive Huffman coding methods are known as the FGK algorithm.
Vitter \cite{Vitter87} proposes an improved technique 
with additional properties and
proves that the number of bits needed in order to encode a message of $n$ characters by his variant, is bounded by the size of the compressed file resulting from the optimal two-pass static Huffman algorithm, {\it plus\/} $n$. In practice, Vitter's method produces often smaller files than static Huffman coding, but not always, and an example for which Vitter's dynamic Huffman coding produces a file that is larger can be found in \cite{KSS}.

An enhanced dynamic Huffman coding named {\it forward looking coding} \cite{KSS} starts with the full frequencies, similar to the static variant, and then decreases them progressively. For this method, after each processed element $\sigma$, the model is altered by {\it decrementing\/} the frequency of $\sigma$ by 1, while the other frequencies remain the same.
Forward looking Huffman coding has been shown to be always better by at least $m-1$ bits than static Huffman coding, where $m$ denotes the size of the alphabet.
We shall refer to the traditional dynamic Huffman coding as {\it backward looking\/}, because its model is based on what has already been seen in the past, unlike the forward looking variant that constructs the model based on what is still to come in the future.
A hybrid method, exploiting both backward and forward approaches is proposed in \cite{FKS}, and 
has been shown to be always at least as good as the forward looking Huffman coding. 

If the model is learned adaptively, as in the traditional backward looking codings, no description of the model is needed, since the model is updated by the encoder and the decoder in synchronization. 
However, in the other mentioned versions, the details of the chosen model on which the method relies, 
are needed for the decoding and 
should be adjoined to the compressed file, for example, as a header.
For static coding, this header may include approximate probabilities or just the set of codeword lengths. However, the forward looking as well as the hybrid variants require the exact frequencies of the elements. When the alphabet is small, the size of the necessary header might often be deemed negligible relative to the size of the input file. Larger alphabets, consisting, e.g., of all the words in a large textual database \cite{hufword}, may often be justified by the fact that the list of different words and their frequencies are needed anyway in an Information Retrieval system.

The contribution of this paper is as follows:
we first define a new generic coding method which we call {\it weighted\/} coding, encompassing all mentioned variants (static, forward and backward) as special cases. 
Second, a new special case called {\it positional\/} is suggested, and shown to be  always at least as good as the forward looking coding.
Third, we present empirical results that show practical improvements of the proposed method, even when the encoded file includes the model description. 

It is important to stress that all the methods can in fact be applied to every 
statistical
coding technique, in particular to arithmetic coding or PPM. This paper brings theoretical and empirical results for both Huffman and arithmetic coding.

The paper is organized as follows. Section~2 introduces the weighted coding concept and shows how certain known compression techniques can be derived from it as special cases. Positional encoding is then  proposed as a new variant, and the proof that it is at least as good as  forward looking coding is given in Section~3.
Section~4 presents empirical results for weighted codings with various parameters, showing its improvement in practice.

\section{Weighted Coding}

\subsection{Definitions}

Given is a file $T=T[1,n]$ of $n$ characters over an alphabet $\Sigma$. We shall define a general weight $W(g,\sigma, \ell, u)$ based on four parameters, in which
\begin{itemize}
    \item[--] $g:[1,n] \longrightarrow {\rm I\!R}^+$ is a non negative function defined on the integers that assigns a positive real number as a weight to each position $i\in [1,n]$ within $T$;
    \item[--] $\sigma\in \Sigma$ is a character of the alphabet; 
    \item[--] $\ell$ and $u$ are the boundaries of an  interval, $1\le \ell\le u\le n$, serving to restrict the domain of the function $g$.
\end{itemize}

The value of the weight $W(g,\sigma, \ell, u)$ will be defined for each character $\sigma \in \Sigma$,   as the sum of the values of the function $g$ for all positions $j$ in the range $[\ell,u]$ at which $\sigma$ occurs, that is $T[j]=\sigma$. Formally
$$ W(g,\sigma, \ell, u)\; = \sum_{\{\ell \leq j \le  u \ \mid \ T[j]=\sigma\}}{g(j)}.$$

 We are in particular interested in two special cases of the weight $W$, defined relatively to a current position $i$,  one we call {\it Backward\/} looking and the other one {\it Forward\/} looking, or backward and forward weights for short. These are implemented by means of the interval $[\ell, u]$ used to restrict the considered range. The {\sl backward weight\/} refers to the positions that have already been processed, i.e.,  
$$W(g,\sigma,1,i-1)\;=\sum_{\{1 \leq j \leq i-1 \ \mid \ T[j]=\sigma\}}{g(j)}
,$$ 
whereas the {\sl forward weight\/} corresponds to the positions yet to come, 
$$W(g,\sigma,i,n)\;= \sum_{\{ {i \le j \le n} \ \mid \ T[j]=\sigma\}}{g(j)}
.$$

\noindent


The aim of this definition is to generalize different existing coding approaches into a consistent framework, so that they can be derived as special cases of weighted coding. This will then lead to the possibility of generating several new variants with improved performances, by varying the parameters to obtain hitherto unknown special cases.

\vspace{3mm}
\noindent
{\sc Static coding} is the special case for which $g$ is the constant function $\mathds{1}\equiv g(i)=1$ for all $i$, and  the weight function, denoted by $W(\mathds{1},\sigma,1,n)$
is constant for all indices. 

\vspace{3mm}
\noindent
The classical {\sc adaptive coding} is a special case of using a backward weight in which, as above,  $g(i)=1$ for all $i$, but unlike {\sc Static coding}, the weights are not constant and are rather recomputed for all indices $1 \le i \le n$ according to backward weights:
\begin{equation*}
W(\mathds{1},\sigma,1,i-1)= \sum_{\{1\le j \le i-1 \ \mid \ T[j]=\sigma\}}{1} = \mbox{number of occurrences of $\sigma$ in } T[1,i-1].
\end{equation*}

\vspace{3mm}
\noindent
{\sc Forward coding} is a special case of using a forward weight in which $g(i)=1$ for all $i$. It is symmetrical to the classical adaptive coding, but computes its model according to suffixes rather than prefixes of the text $T$.
That is, 
\begin{equation*}
W(\mathds{1},\sigma,i,n)= \sum_{\{i\le j \le n \ \mid \ T[j]=\sigma\}}{1} = \mbox{number of occurrences of $\sigma$ in } T[i,n].
\end{equation*}

The idea behind the  extension below is the following. {\sc Static} Huffman encodes a character $\sigma$ by the same codeword ${\cal E}(\sigma)$, regardless of where in the text $\sigma$ occurs. The choice of how many bits to allocate to ${\cal E}(\sigma)$ is therefore governed solely by the frequency of $\sigma$ in $T$, and not by where in $T$ the occurrences of $\sigma$ can be found. In the {\sc Adaptive} approach, on the other hand, the set of frequencies in the entire file $T$ are yet unknown after only a prefix of size $i-1$ has been processed, for $i\le n$. Basing the encoding then on the currently known statistics is thus just an {\it estimate\/}, and the good performance of such an approach depends on whether or not the distribution of the characters derived from the processed prefix is similar to the distribution in the entire file. Backward adaptive methods take advantage of the fact that the damage of using wrong frequencies is limited, since the numbers and thus also the corresponding codewords are constantly updated, and the latter will ultimately achieve their optimal lengths.
By reversing the process to consider the future rather than the past, the {\sc Forward} coding again deals with correct frequencies, and not just with estimates. 

However, once the psychological barrier forcing us to base our encoding models on frequencies or their estimates has been broken,
it might be justified to deviate from the common practice and try a {\it greedy\/} approach for a more convenient definition of the model. 
In particular, characters that are close to the current position might be assigned a higher priority than those farther away. The rationale of such an assignment is that the close by characters are those that we are about to encode, so we concentrate on how to reduce the lengths of their codewords, even at the price of having to lengthen the codewords of more distant characters in the text, since, anyway, the encoding of those will be reconsidered by the adaptive process once we get closer to them. 

The assignment of differing priorities or {\it weights\/} can be materialized by using a decreasing function $g$ instead of a constant one. The simplest option would be a linear decrease, which leads to the following definition.

\bigskip
\noindent
{\sc Positional coding}, first defined in this paper, is a special case of a forward weight, with ${\cal L}\equiv g(i)=n-i+1$ for $1 \leq i \leq n$, where $n=|T|$. We shall use the notation $p_\sigma(i)$ to denote $W({\cal L},\sigma,i,n)$.

\medskip


Note that the idea of giving increased attention to closer rather than to more distant elements is not new to data compression. A similar choice appears when choosing a sliding window of limited size in Ziv-Lempel coding \cite{LZSS}, and when the accumulated frequencies are periodically rescaled in adaptive (backward) Huffman or arithmetic coding, as suggested in \cite{Nelson}. 

As we shall see in the experimental section below, the intuition of assigning higher weights to closer elements pays off and indeed yields improvements in the compression performance. This leads naturally to pushing the idea even further. The extreme case would be an exponentially decreasing function $g$, e.g., ${\cal E}\equiv g(i)=2^{n-i}$. In this case, the weight of the following character to be processed will always be the largest, 
since even when the suffix of the text is of the form {\tt abbb$\cdots$b}, we get that 
$W({\cal E},{\tt a},i,n)=2^{n-i}>\sum_{j=i+1}^n 2^{n-j}=
W({\cal E},{\tt b},i,n)$.
It follows that the codeword assigned by Huffman's algorithm at position $i$ will be of length 1 bit. Therefore, using this exponential function $\cal E$ for $g$, the text will be encoded by exactly $n$ bits, one bit per character, which means that the  bulk of the information is encoded in the header. 

This encoding of the header is, however, very costly. The weight of each of the characters may be of the order of $2^n$, requiring $\theta(n)$ bits for its encoding, so the header may be of size  $O(|\Sigma| n)$. Moreover, the update algorithms will be very time consuming, having to deal with numbers of unbounded precision. The challenge is therefore to find reasonable functions $g$, which yield a good tradeoff between encoding the text and the header, and we aim at minimizing the sum of their sizes.

\subsection{Detailed comparative example}

To clarify these definition, we present the text $T=$ {\tt c c a b b b c a a a}$\;$ as a small running example and compare  its different encodings.
Although it could, in principle, be illustrated also by arithmetic coding variants, we restrict the example to Huffman encoding alternatives which are easier to follow. 

\subsubsection{Positional coding}

Recall that we use $p_\sigma(i)$ to denote $W({\cal L},\sigma,i,n)$.
The details for the Positional encoding are presented in Table~\ref{PositionalExample}.
The function ${\cal L}$, given on the second line, enumerates the indices in reverse order starting at $n=10$ down to 1.
At the first position $i=1$, the values of $p_{\tt a}(1)$, 
$p_{\tt b}(1)$ and $p_{\tt c}(1)$ are 14, 18 and 23, respectively, as shown in the first column of the last three rows. 
In a left to right scan, the values $p_\sigma(i)$ only change at indices $i$ for which
$T[i-1]=\sigma$. Light gray therefore refers to the non-changed values (starting, in a left to right scan, just after an occurrence of $\sigma$ and ending at the rightmost position where $\sigma$ occurs in $T$).

\renewcommand{\arraystretch}{1.5}

\begin{table}
\begin{center}
\begin{tabular}{c|c|c|c|c|c|c|c|c|c|c|}
\multicolumn{1}{c}{$i$}&\multicolumn{1}{c}{\makebox[4mm]{1}}& \multicolumn{1}{c}{\makebox[4mm]{2}}& 
\multicolumn{1}{c}{\makebox[4mm]{3}}& 
\multicolumn{1}{c}{\makebox[4mm]{4}}& 
\multicolumn{1}{c}{\makebox[4mm]{5}}& 
\multicolumn{1}{c}{\makebox[4mm]{6}}& 
\multicolumn{1}{c}{\makebox[4mm]{7}}&
\multicolumn{1}{c}{\makebox[4mm]{8}}&
\multicolumn{1}{c}{\makebox[4mm]{9}}&
\multicolumn{1}{c}{\makebox[4mm]{10}}\\[2mm] \cline{2-11}
$T$ & {\tt c} & {\tt c} & {\tt a} & {\tt b} & {\tt b} & {\tt b} & {\tt c} & {\tt a} & {\tt a} & {\tt a} \\ \cline{2-11}
${\cal L}(i)$ & 10 & 9 & 8 & 7 & 6 & 5 & 4 & 3 & 2 & 1 \\ \cline{2-11}
$p_{\tt a}(i)$ & 14 & {\color{mygray} 14} & {\color{mygray} 14} &  6 & {\color{mygray} 6} & {\color{mygray} 6} & {\color{mygray} 6} & {\color{mygray} 6} & 3 & 1 \\
\cline{2-11}
$p_{\tt b}(i)$ & 18 & {\color{mygray} 18} & {\color{mygray} 18} & {\color{mygray} 18} & 11 & 5 & 0 & {\color{mygray} 0} & {\color{mygray} 0} &  {\color{mygray} 0}\\ 
\cline{2-11}
$p_{\tt c}(i)$ 
& 23 & 13 & 4 & {\color{mygray} 4} & {\color{mygray} 4} & {\color{mygray} 4} & {\color{mygray} 4} & 0 & {\color{mygray} 0} & {\color{mygray} 0} \\ 
\cline{2-11}
\end{tabular}
\end{center}
\begin{center}
\caption{\small\sl Positional Coding example for $T={\tt ccabbbcaaa}$.}
\label{PositionalExample}
\end{center}
\end{table}

\begin{figure}[h]
\begin{center}
\includegraphics[scale=0.12]{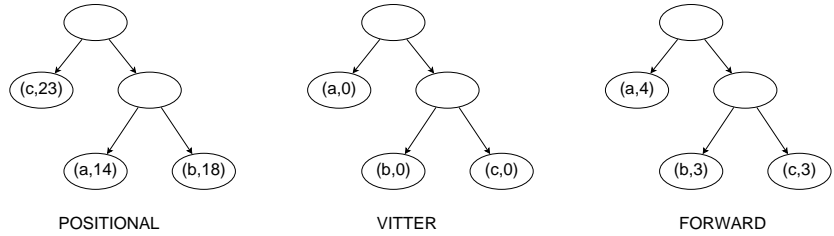}
\end{center}
\caption{\small\sl 
Initial trees for $T=$ {\tt  ccabbbcaaa}.}
\label{Trees}
\end{figure}

The Huffman tree is initialized with the weights of position 1, as shown in the left part of Figure~\ref{Trees}. The first {\tt c} is encoded by 0, and its weight is then decremented by ${\cal L}(1)=10$ from 23 to 13. 
The tree gets updated, now having leaves with weights 14, 18 and 13 for {\tt a}, {\tt b} and 
{\tt c}, respectively, as given in column 2 of Table~\ref{PositionalExample}. The second {\tt c} is therefore encoded by the two bits 10.
The weight of {\tt c} is decremented by ${\cal L}(2)=9$ from 13 to 4, and the following character {\tt a} is encoded by the two bits 11. The weights are then updated to
6, 18 and 4, as shown in column 4 of Table~\ref{PositionalExample}. At this stage, the character {\tt b} has become the one with the shortest codeword, and the two following {\tt b}s are encoded each by 0,
updating the weight of {\tt b} first to 11 and then to 5, so that the encoding for the last {\tt b} becomes 11.  
After the last {\tt b} is processed, it is removed from the tree as its frequency has become 0, resulting in a tree containing only {\tt c} and {\tt a}.
When the last {\tt c} is processed, the codeword 0 is output, the leaf for {\tt c} is removed from the tree, and the tree remains with 
a single node corresponding to {\tt a}.
Since the decoder also discovers that the alphabet of the remaining suffix of the file  contains only a single character, which must be {\tt a}, with weight  $p_a(8) = \sum_{i=8}^{10}{{\cal L}(i)}=6$, the number of its repetitions can be calculated, thus no additional bits need to be transferred.

\subsubsection{Vitter's backward coding}

The model in Vitter's algorithm need not be transmitted to the decoder as it is learnt incrementally while processing the encoded file.
We assume that the exact alphabet is known to both encoder and decoder, and do not use the special {\it Not-Yet-Transmitted\/} 
leaf suggested by Vitter for dealing with newly encountered symbols.
The initial tree for Vitter's algorithm is 
the middle one in Figure~\ref{Trees}, starting with all frequencies equal to 0. 

The correctness of Vitter's algorithm relies on the {\it sibling property\/} \cite{Gallager}, which states that a tree is a Huffman tree if and only if its nodes can be listed by nonincreasing weight so that each node is adjacent to its sibling. We shall use the convention to place the nodes in a Huffman tree in such a way that this list can be obtained by a bottom-up, left to right scan of the nodes. 

The codeword for {\tt c} is 11, and its frequency is incremented by 1, resulting in a shorter codeword, 0, for the following {\tt c}, by swapping the leaves of {\tt a} and {\tt c}.
The next character {\tt a} is encoded by 11, and its frequency is updated, but the tree remains unchanged.
The following two characters are {\tt b} and {\tt b}, both of which are encoded by 10. The frequency of {\tt b} gets updated first to 1, then to 2, resulting in a swap with the leaf for {\tt a} in order to retain the sibling property. 
For the next and last {\tt b} the codeword 11 is output, its frequency is increased to 3, and its leaf is swapped with the leaf corresponding to {\tt c}. 
The following character {\tt c} results in the output of the codeword 11 and an update of its frequency to 3. 
The codeword for each of the last three occurrences of {\tt a} is then 10, incrementing the weight of {\tt a} from 1 to 2, 3, and 4, where the tree is changed only at the last step, only after the last {\tt a} has already been encoded.

\subsubsection{Forward coding}

The {\sc forward} coding  algorithm basically works in the opposite way, starting with the final tree of the dynamic variant, and ending with the empty tree. 
The tree is initialized with weights 
($W(\mathds{1},{\tt a},1,n)$, $W(\mathds{1},{\tt b},1,n)$, $W(\mathds{1},{\tt c},1,n))
=(4,3,3)$, like for static Huffman encoding, as shown in the right part of  Figure~\ref{Trees}. The first {\tt c} is encoded by 11, and the frequency of {\tt c} is decremented to 2, resulting in an interchange of {\tt b} and {\tt c}. 
The second {\tt c} is therefore encoded by 10, and its frequency is updated to 1.
The character {\tt a} is then encoded by 0 and its weight is decremented to 3. The following three {\tt b}s are encoded by 11 and the frequency for {\tt b} is repeatedly decremented to 0 and its leaf is finally removed from the tree. The tree remains with 2 leaves for {\tt a} and {\tt c}, and the following character {\tt c} is encoded by 0. The last three {\tt a}s need not be encoded as for positional coding.

\setlength{\tabcolsep}{0.5em}
\begin{figure}[!h]
\small
\begin{center}
\begin{tabular}{@{\hspace{0mm}}l@{\hspace{6mm}}ccc|cccccccccc@{\hspace{0mm}}}
& \multicolumn{3}{l|}{Weights}\\[-2.5mm] 
& {\tt a} & {\tt b} & {\tt c} & \raisebox{3mm}{\tt c} & \raisebox{3mm}{\tt c} & \raisebox{3mm}{\tt a} & \raisebox{3mm}{\tt b} & \raisebox{3mm}{\tt b} & \raisebox{3mm}{\tt b} & \raisebox{3mm}{\tt c} & \raisebox{3mm}{\tt a} & \raisebox{3mm}{\tt a} & \raisebox{3mm}{\tt a}\\ \hline
{\sc Positional} & {\tiny 14} & {\tiny 18} & {\tiny 23} & 0 & 10 & 11 & 0 & 0 & 11 & 0 & -- & -- & -- \\
{\sc Vitter} & -- & -- & -- & 11 & 0 & 11 & 10 & 10 & 11 & 11 & 10 & 10 & 10 \\
{\sc Forward} & {\tiny 4} & {\tiny 3} & {\tiny 3} & 11 & 10 & 0 & 11 & 11 & 11 & 0 & -- & -- & --\\
\hline
\end{tabular}
\vspace{3mm}

\caption{\small\sl The encoding of $T={\tt ccabbbcaaa}$ using the three adaptive techniques.}
\end{center}
\label{Coding Example}
\end{figure}

Figure~2 summarizes the different encodings of this example when applying Huffman to these weights. The first three columns of the table display the initial weights of the symbols; these should be encoded and prepended to the compressed file. The rest of the table shows the binary output sequences produced by the different approaches. 
Although all the reported outcomes of our experimental results include the appropriate header for each method (this header is empty for Vitter's algorithm), we do not include any precise binary encoding in this small example; it is generally of secondary importance relative to the size of real life input files, but it might distort the outcome 
of the comparison on a small artificial example as this one.

The net number of bits required to encode $T$ for this example by the three alternatives is
10, 19 and 12 for {\sc Positional}, {\sc Vitter} and {\sc Forward}, respectively.
Note that the first {\tt a} is encoded by a single bit by {\sc Forward} and by two bits by {\sc Positional}, illustrating that although there are overall savings in space, the individual codewords assigned by {\sc Forward} may be locally shorter than the corresponding ones of {\sc Positional}.

\section{Analysis}

This section provides a proof showing that {\sc positional} coding is at least as good as {\sc forward}, which in turn has been proven to be always better than {\sc static} Huffman coding by at least $m-1$ bits, and better than Vitter's {\sc dynamic} algorithm in the worst case.
We use the phrase $A$ {\sf is at least as good as} $B$ to assert that the size of the encoded file according to method $A$ is not larger than the size of this file according to method $B$.

One of the conspicuous features of weighted coding in general, and of {\sc Positional} coding in particular, is the fact that while the true distribution of the characters in $\Sigma$ is known, it is a {\it different\/} distribution that is used as basis to derive the encoding. We need a measure to quantify the loss (or gain) in compression efficiency incurred by this change of strategy. 

If arithmetic coding is used, the average expected codeword length for an assumed true probability distribution $P=(p_1,\ldots,p_m)$ is  $-\sum_{i=1}^m p_i\log p_i$, the entropy of $P$. If another distribution $Q=(q_1,\ldots,q_m)$ is used instead of $P$, the average expected codeword length will be  $-\sum_{i=1}^m p_i\log q_i$, and their difference, known as the Kullback-Leibler divergence \cite{KL}
$$D_{\mbox{\scriptsize KL}}(P\,\|\,Q)= -\sum_{i=1}^m p_i\log q_i -\Big(-\sum_{i=1}^m p_i\log p_i\Big)=\sum_{i=1}^m p_i\log\left(\frac{p_i}{q_i}\right),$$
is well known to be non-negative, as shown by Gibb's inequality. We shall use this measure to show that the encoding according to the {\sc Positional} scheme is at least as good as using a {\sc Static} encoding. This will be exact for arithmetic coding and only an approximation for Huffman coding. The accuracy of the approximation depends on how close the distribution is to a {\it dyadic\/} one, in which all the probabilities are powers of $\frac{1}{2}$. Indeed, on dyadic distributions, Huffman and arithmetic coding yield identical average codeword lengths, or even completely identical codes in certain cases \cite{KSRandomness}, and a bound for the deviation of Huffman from arithmetic coding can be found in \cite{LG}.

\vspace*{3mm}
\noindent {\sf Lemma}~1:\quad  {\sl 
 Given an index $t$, $1 \leq t < n$, then encoding $T[1,t]$ using {\sc Forward} for the first $t$ characters, that is, with  weights 
 $W(\mathds{1},\sigma,i,t)$, 
 for varying values of $i$ from 1 to $t$,
 followed by using {\sc Forward} for the last $n-t$ characters of $T$ with weights 
 $W(\mathds{1},\sigma,i,n)$,
 for varying values of $i$ from $t+1$ to $n$,
 is at least as good as using {\sc Forward} for the entire message, 
 i.e.,  using weights 
 $W(\mathds{1},\sigma,i,n)$, $1\le i \le n$.}

\vspace*{3mm}
\noindent {\sf Proof}:\quad  The difference between the encodings of applying {\sc Forward} to the entire message, versus applying it to the first and second parts separately, is the encoding of the prefix of $T$ up to position $t$. 
Splitting the encoding into two parts considers, while encoding the first part, only the number of occurrences of any character $\sigma$ within the first $t$ characters,
rather than in the entire text of $n$ characters. The latter also takes into account the occurrences of $\sigma$ in the suffix of $T$ of size $n-t$, which are not relevant for encoding just the prefix of size $t$.
The resulting encoding using the split into two parts can therefore not be worse, 
because {\sc Forward} Huffman and {\sc Forward} arithmetic coding are better than {\sc Static}, which is known to be optimal if the true distribution is used.\hfill\qed

\vspace*{3mm}
We actually need a generalization of Lemma~1 to work for {\sc Weighted} adaptive coding and not only for the special case of {\sc Forward}, as defined in the following lemma.

\vspace*{3mm}
\noindent {\sf Lemma}~2:\quad  {\sl 
Given an index $t$, $1 \leq t < n$, a non negative function $g:[1,n] \longrightarrow {\rm I\!R}^+$, and two real parameters $0 \leq x_1 \leq x_2$, then encoding $T[1,t]$ with  weights 
 $$W(\mathds{1},\sigma,i,t)+x_1 \cdot W(g,\sigma,t+1,n), \quad 
 \mbox{for }i=1,\ldots,t
 \eqno(1)$$%
(taking into account also occurrences of $\sigma$ in the suffix $T[t+1,n]$ of $T$), is at least as good as encoding 
 $T[1,t]$ with  weights 
 $$W(\mathds{1},\sigma,i,t)+x_2 \cdot W(g,\sigma,t+1,n)\quad \mbox{for }i=1,\ldots,t.\eqno(2)$$}
 
Note that we concentrate on encoding the prefix of $T$ up to position $t$. 
The first summand 
$W(\mathds{1},\sigma,i,t)$
computes
the number of occurrences of any character $\sigma$ 
within $[i,t]$ according to {\sc Forward} that
is better than {\sc Static}; the latter encoding is known to be an optimal static encoding
for this true distribution. The second summand, however,
$x_i \cdot W(g,\sigma,t+1,n)$,
considers also the number of occurrences of $\sigma$ in the suffix of $T$ of size $n-t$ according to a function $g$, and multiplies this weight by some real constant $x_1$ or $x_2$, so that the resulting weights deviate from the true distribution. We want to show that this deviation increases with  larger weights in the second summand. 

Lemma~2 generalizes Lemma~1, which corresponds to the particular choice of the parameters  $x_1=0$, $x_2=1$, and function $g=\mathds{1}$. To prove Lemma~2, we need the following technical claims.

Lemma 3 shows how to translate the change of weights of Lemma~2 into a change of probabilities.

\vspace*{3mm}

\noindent{\sf Lemma}~3:\quad 
{\sl
Let $P$ be the probability distribution of character occurrences corresponding to the weights 
$W(\mathds{1},\sigma,i,t)$, for $ i=1,\ldots,t$,
and let $Q$ and $Q'$ be the probability distributions corresponding to the weights defined by equations (1) and (2), respectively.
For every character $\sigma$ let $p_\sigma$, $q_\sigma$ and $q'_\sigma$ be the probability of $\sigma$ in distributions $P$, $Q$ and $Q'$, respectively. 
Then either 

$$p_\sigma \le q_\sigma \le q'_\sigma \qquad \mbox{or}\qquad p_\sigma \ge q_\sigma \ge q'_\sigma,$$
that is, $Q$ is closer to $P$ than $Q'$.
}

\vspace{3mm}

\noindent{\sf Proof:} 
Given an index $t$, $1 \leq t < n$ and fixed indices $i$, $1\le i \le t$, we compute the probability  $p_\sigma(x)$ of $\sigma$,
based on the number of its occurrences to the left and to the right of $t$, as a function of $x$.
$$p_\sigma(x)=  \frac{\displaystyle
    W(\mathds{1},\sigma,i,t)+x\cdot W(g,\sigma,t+1,n)}{\displaystyle
    \sum_{\sigma \in \Sigma} W(\mathds{1},\sigma,i,t)+x\cdot \sum_{\sigma \in \Sigma} W(g,\sigma,t+1,n)}\equiv \frac{A+x\;B}{C + x\;D}.$$

Its derivative relative to $x$ is   
    
$$p_\sigma'(x)=  \frac{BC-AD}{(C+x\;D)^2}=\frac{CD\big(\frac{B}{D}-\frac{A}{C} \big)}{(C+x\;D)^2}.$$
The sign of $p_\sigma'(x)$  depends on its numerator, which in turn depends on the relation between 
$$ \frac{B}{D}=\frac{\displaystyle W(g,\sigma,t+1,n)}{\displaystyle \sum_{\sigma \in \Sigma}W(g,\sigma,t+1,n)}\qquad \mbox{and}\qquad \frac{A}{C}=\frac{\displaystyle W(\mathds{1},\sigma,i,t)}{\displaystyle \sum_{\sigma \in \Sigma}W(\mathds{1},\sigma,i,t)} .$$

That is, the sign of $p_\sigma'(x)$ depends, for each index $i$ and each  character $\sigma$, on whether the probability of $\sigma$ 
in $[i,t]$
is smaller or larger than the probability according to the function $g$ on the right of $t$. In particular, this fact is independent of $x$. Therefore, as a function of $x$, the  probabilities are either increasing or decreasing for each $\sigma$. \hfill\qed

\vspace*{3mm}
The next lemma shows that it is possible to bridge the gap between different probability distributions by smaller steps in each of which only two probabilities are changed.

\noindent {\sf Lemma}~4:\quad  {\sl Let $P=(p_1,\ldots,p_m)$ and $Q=(q_1,\ldots,q_m)$ be two different probability distributions. 
Then it is possible to define a sequence of probability distributions $R_0, R_1,\ldots,R_k$, with $R_0=P$ and $R_k=Q$, such that $R_j$ and $R_{j+1}$ differ only in two of their $m$ coordinates, for $j=0,\ldots, k-1$. 
}

\vspace*{3mm}
\noindent {\sf Proof}:\quad  Denote by  $z \le m$ the number of indices for which $p_i\neq q_i$. 
We construct a sequence  of $k \le z$ probability distributions.
It suffices to show, for each $j\ge 0$, that given $R_j$, which differs from $Q$ in $h\le z$ coordinates, it is possible to construct a distribution $R_{j+1}$, which differs from $Q$ in strictly less than $h$ coordinates.
Denote $R_j=(u_1,\ldots,u_m)$ and $R_{j+1}=(v_1,\ldots,v_m)$, we need to show that there exist indices $1\le a,b\le m$ and a non-zero real number $\epsilon_j\neq0$ such that
$$v_a=u_a-\epsilon_j\qquad v_b=u_b+\epsilon_j\qquad \mbox{and}\quad v_c=u_c\mbox{ for }c\neq a,b.\eqno(3)$$

Consider the differences $d_i=u_i-q_i$ for all $1\le i\le m$ and let $a$ be an index for which $|d_i|$ is minimal while still being non-zero; $d_i$ can be positive or negative. Define then
$R_{j+1}=(v_1,\ldots,v_m)$ as given in eq.~(3), with $\epsilon_j=d_a$. We have to show that $R_{j+1}$ is indeed a probability distribution. Clearly $\sum_{i=1}^m v_i=\sum_{i=1}^m u_i=1$. We have
$$v_a=u_a-d_a= q_a,$$
so that $v_a$ is a probability. There were $h$ coordinates in $R_j$ differing from those of $Q$, including the index $a$, but now $v_a=q_a$, so the number of differing coordinates is at most $h-1$ (actually, it is either $h-1$ or $h-2$ in the special case that $v_b=u_b+d_a$ is equal to $q_b$). We still need to show that $0\le v_b\le 1$. The proof depends on whether $d_a$ is positive or negative.
\begin{enumerate}
    \item If $d_a>0$, then $v_b$ is larger than $u_b$, so it is clearly positive. On the other hand
    $u_b\le 1-u_a$ and thus
    $$v_b=u_b+d_a\le (1-u_a) +(u_a-q_a) =1-q_a\le 1.$$
    \item If $d_a<0$, then $v_b$ is smaller than $u_b$, so it is clearly smaller than 1. On the other hand, there must be an index $b$ for which $d_b>0$. Moreover, since $a$ was chosen such that $|d_a|$ is minimal, $|d_b|\ge |d_a|$, so that $u_b-q_b\ge q_a-u_a$. Therefore
    $$v_b=u_b+d_a\ge (q_a-u_a+q_b) + (u_a-q_a) = q_b\ge 0.\eqno\qed$$
\end{enumerate}

\vspace{5mm}

The following lemma shows that constructing the distribution $R_{j+1}$ from $R_j$ 
forms a monotonic sequence of probabilities for each character $\sigma$.
More precisely, there are $m$ sequences of increasing or decreasing probabilities.

\vspace*{3mm}

\noindent{\sf Lemma}~5:\quad 
{\sl
Let $R_j$, $0 \le j \le k$, be the probability distributions constructed by the proof of Lemma~4, and let $r_j(\sigma)$ denote the probability of $\sigma$ in distribution $R_j$.
Then  the sequence $\{r_j(\sigma)\}_{j=0}^k$ is monotonic for each $\sigma$.
}

\vspace{3mm}

\vspace{3mm}

\noindent{\sf Proof:} 
Set $\sigma \in \Sigma$.
It is sufficient to show that for every $j$, the three elements $r_j$, $r_{j+1}$ and $r_k$ form a  monotonic sequence. 
Following the construction of $R_j$,
the distributions $R_{j}$ and $R_{j+1}$ are identical, except for two indices $1\le a,b\le m$, where $a$ is the index for which the difference $d$ between a coordinate of $R_j$ and the corresponding coordinate of $Q=R_k$ has  minimal absolute value,
while still being non-zero. 
That is, 

\vspace*{-3mm}
$$ d=r_k(a)-r_j(a)\qquad\mbox{and}\qquad |d|\le |r_k(b)-r_j(b)|\quad\mbox{for all $b$}\qquad \mbox{and}$$

\vspace*{-3mm}
$$r_{j+1}(a)=r_j(a)+d\qquad \mbox{and}\qquad   r_{j+1}(b)=r_j(b)-d.$$
Note that once $r_j(\sigma)$ is set to $r_k(\sigma)$, it never changes.
Moreover, since all\linebreak $\{r_i(\sigma)\}_{\sigma\in\Sigma}$ are probability distributions,
it follows that 
if $r_k(a)-r_j(a)=d>0$ then for the index $b$, it holds that
$r_k(b)-r_j(b)<0$, and vice versa. 


\vspace{3mm}
There are two cases:
\begin{enumerate}
\item 
If $d>0$, then $d\leq r_j(b)-r_k(b)$, and 
$r_j(a) < r_j(a)+d=r_{j+1}(a)=r_k(a)$,
is non-decreasing for $a$.
In addition, 
$r_k(b)\leq r_j(b)-d=r_{j+1}(b)< r_j(b)$, as $d>0$, 
so $r_j(b)>r_{j+1}(b)\geq r_t(b)$
is non-increasing for $b$.\\[-1mm]
\item
If $d<0$, then $d\geq r_j(b)-r_k(b)$, and 
$r_j(a) > r_j(a)+d=r_{j+1}(a)=r_k(a)$,
is non-increasing for $a$.
In addition, 
$r_k(b)\geq r_j(b)-d=r_{j+1}(b)> r_j(b)$, 
so $r_j(b)<r_{j+1}(b)\leq r_k(b)$
is non-decreasing for $b$.\hfill\qed
\end{enumerate}

The following lemma shows that the size of the encoded file is an increasing function of the size of the deviation from the true probability vector for the case that only two of its probabilities are altered.

\vspace*{3mm}
\noindent {\sf Lemma}~6:\quad  {\sl Let $P=(p_1,\ldots,p_m)$ and $Q=(q_1,\ldots,q_m)$ be two different probability distributions, differing only in two coordinates $1\le a,b\le m$, that is, there exists a real number $x>0$, such that
$$q_a = p_a -x\qquad\mbox{and}\qquad q_b = p_b +x.$$
Then $D_{\mbox{\scriptsize KL}}(P\,\|\,Q)$, representing the increase in the average codeword length caused by substituting $Q$ to $P$, is an increasing function of $x$.
}

\vspace*{3mm}
\noindent {\sf Proof}:\quad  Define the function $f(x)=D_{\mbox{\scriptsize KL}}(P\,\|\,Q)$ using the above notations and consider its derivative $f'(x)$. Recall that all the probabilities $p_i$ and $q_i$ are constants relative to $x$, so we have
$$f(x)= p_a\,\log \frac{p_a}{p_a-x} \;\;+\;\; p_b\,\log \frac{p_b}{p_b+x}.$$
Therefore
$$f'(x) = \frac{1}{\ln 2}\left(\frac{p_a}{p_a-x}\;\;-\;\; \frac{p_b}{p_b+x}\right) \;\;\ge\;\;0,$$
because the first term in the parentheses is larger than 1 while the second is strictly smaller than 1.\hfill\qed

\vspace*{3mm}
\noindent{\sf Proof of Lemma 2:} Let $P$  be the  probability distribution of character occurrences corresponding to the  weights
$W(\mathds{1},\sigma,i,t)$, for $i=1,\ldots,t$,
that is, the true distribution in $[i,t]$, and let $Q$ be the distribution  corresponding to the  weights 
$W(\mathds{1},\sigma,i,t)+x \cdot W(g,\sigma,t+1,n)$, for $i=1,\ldots,t$,
defined in equations (1) and (2) for different values of $x$.

Let $R_0,\ldots,R_k$ be the sequence of probability distributions defined accordingly in Lemma~4. 
By Lemma~3 and Lemma~5, the increased weights of equation~(2) relative to equation~(1) correspond to probability distributions in which each coordinate increases its absolute distance from the corresponding coordinate in the true probability distribution.

By Lemma~6 we know that the increase in the average codeword length caused by the passage from  $R_j$ to $R_{j+1}$ is an increasing function of the difference of the only probabilities that change in this passage, for all $0\le j<k$, so that the increase in the average codeword length caused by the passage from  $R_0=P$ to $R_{k}=Q$ is an increasing function of the changing probabilities, which all depend on the parameter $x$. Thus larger values of $x$ imply a larger increase.\hfill\qed

\vspace*{5mm}
\noindent {\sf Theorem}:\quad  {\sl For a given file $T$ of length $n$, the average codeword length of {\sc Positional} is at least as good as the average codeword length of {\sc Forward} coding.
}

\vspace*{3mm}
\noindent{\sf Proof:} We construct a sequence of functions ${\cal G}=\{g_j\}_{j=1}^n$ as follows, the first one $g_1$ being the function corresponding to {\sc Forward} and the last one $g_n$ to {\sc Positional}.
The function $g_1$ is thus the constant function $\mathds{1}\equiv g_1(i)=1$. 
For $j \ge 2$, we define $g_j$ recursively by:
\begin{equation*}
  g_j(i) = \begin{cases}
        j &\qquad {i \leq (n-j+1)}\\
        g_{j-1}(i)  & \qquad {i > (n-j+1)}, \\
        \end{cases}
  \label{eqg}
\end{equation*}
so $g_j$ is constant up to $n-j+1$ and then decreases linearly, see Table~2 for an illustration.

\vspace*{4mm}
\begin{minipage}[b]{0.95\linewidth}
\centering
{\renewcommand{\arraystretch}{1.5}
\centering
\scriptsize
\begin{tabular}{ccccccccccc}
\multicolumn{1}{c}{}&\multicolumn{1}{c}{1}& \multicolumn{1}{c}{2}& \multicolumn{1}{c}{3}& 
\multicolumn{1}{c}{4}& 
\multicolumn{1}{c}{5}& 
\multicolumn{1}{c}{6}& 
\multicolumn{1}{c}{7}&
\multicolumn{1}{c}{8}&
\multicolumn{1}{c}{9}&
\multicolumn{1}{c}{10}\\ \cline{2-11}
$T$ & {\tt c} & {\tt c} & {\tt a} & {\tt b} & {\tt b} & {\tt b} & {\tt c} & {\tt a} & {\tt a} & {\tt a} \\ \cline{2-11}
$g_1(i)$ & 1 & 1 & 1 & 1 & 1 & 1 & 1 & 1 & 1 & 1 \\ 
$g_2(i)$ & 2 & 2 & 2 & 2 & 2 & 2 & 2 & 2 & 2 & 1 \\ 
$g_3(i)$ & 3 & 3 & 3 & 3 & 3 & 3 & 3 & 3 & 2 & 1 \\ 
$g_4(i)$ & 4 & 4 & 4 & 4 & 4 & 4 & 4 & 3 & 2 & 1 \\[-1mm]
&\multicolumn{10}{c}{$\cdots$} \\ \cline{2-11}
${\cal L}(i)$ & 10 & 9 & 8 & 7 & 6 & 5 & 4 & 3 & 2 & 1 \\ \cline{2-11} 
\cline{2-11}
\end{tabular}
}
\vspace*{2mm}
\begin{center}
{\small{\sc Table 2:} {\sl Positional Coding example.}}
\end{center}
\vspace*{2mm}
\end{minipage}







We show that the encoding based on $g_{j+1}$ is at least as good as 
the encoding based on $g_j$, for all $j$, so that ultimately, {\sc Positional} is at least as good as {\sc Forward}.

Consider first the suffix $T[n-j+1,n]$ of $T$. 
For all characters $\sigma \in \Sigma$, the weights
$W(g_j,\sigma,i,n)$ and $W(g_{j+1},\sigma,i,n)$, 
for $i$ varying from $n-j+1$ to $n$,
are identical for both functions, as visualized in Figure~3,  so the encoding of the suffix $T[n-j+1,n]$ will be the same for $g_j$ and for $g_{j+1}$.  

\vspace*{4mm}
\setlength{\tabcolsep}{0.5em}
\begin{figure}[!h]
\small
\begin{center}
\scriptsize
\begin{tabular}{@{\hspace{0mm}}c@{\hspace{6mm}}cccc|c|c|c|cccccc@{\hspace{0mm}}}
& 1 & 2 & 3 & $\cdots$ & $n\!-\!j\!-\!1$ & $n\!-\!j$ & $n\!-\!j\!+\!1$ & & $\cdots$ & $n\!-\!2$ & $n\!-\!1$ & $n$\\
\cline{2-13}
\multicolumn{13}{c}{}\\[-2mm]
$g_j$ & $j$ & $j$ & $j$ & & $j$ & $j$ & $j$ & $j\!-\!1$ & &  3 & 2 & 1\\
$g_{j+1}$ & $j\!+\!1$ & $j\!+\!1$ & $j\!+\!1$ & & $j\!+\!1$ & $j\!+\!1$ & $j$ & $j\!-\!1$ & &  3 & 2 & 1\\
\multicolumn{13}{c}{}\\[-5mm]

\end{tabular}
\vspace{3mm}

\caption{\small\sl Schematic view of function $g_j$ and $g_{j+1}$.}
\label{partition}
\end{center}
\end{figure}

For encoding the first $n-j$ positions, we consider the weights for the entire interval $[1,n]$ based on functions $g_j$ and $g_{j+1}$, as illustrated in Figure~\ref{partition},
and rewrite the weights 
as follows: for each $i$ in the range  
$1\leq i \leq n-j$,
$$W(g_j,\sigma,i,n) = j \cdot \Big( W(\mathds{1},\sigma,i,n-j)+\frac{1}{j} \cdot W(g_j,\sigma,n-j+1,n) \Big),\quad\mbox{and}$$
$$W(g_{j+1},\sigma,i,n) = (j+1) \cdot \Big( W(\mathds{1},\sigma,i,n-j)+\frac{1}{j+1} \cdot W(g_j,\sigma,n-j+1,n) \Big),$$
because the weights $W$ for $g_j$ and $g_{j+1}$
are the same for the suffix starting at $n-j+1$. 
Since $\frac{1}{j+1}<\frac{1}{j}$, we can apply Lemma~2 and get that 
encoding $T[1,n-j]$ with  weights 
$W(\mathds{1},\sigma,i,n-j)+\frac{1}{j+1} \cdot W(g_j,\sigma,n-j+1,n)$, for $1\le i \le n-j$,
 is at least as good as encoding $T[1,n-j]$ with  weights 
 $W(\mathds{1},\sigma,i,n-j)+\frac{1}{j} \cdot W(g_j,\sigma,n-j+1,n)$,
 for $1\le i \le n-j$.
 The multiplication by $j$ and $j+1$, respectively, changes the absolute weights, but preserves the {\it relative\/} weights for all $\sigma \in \Sigma$, and thus, the corresponding encodings.
 
Combining the encodings for the prefix $T[1,n-j]$ of $T$ and the suffix $T[n-j+1,n]$ of $T$ 
concludes the proof.\hfill\qed





\begin{figure}[!h]
\begin{center}

\pgfplotsset{
  compat=newest,
  width=12cm,
  height=5cm,
  every tick label/.append style = {font=\tiny},
  every axis label/.append style = {font=\scriptsize}
}

\begin{pycode}
from collections import defaultdict
data1 = defaultdict(list)
with open('mydata2.dat') as dd:
    for line in dd:
    	if line.startswith('='):
            if 'k' in line:
                data1['k'] += [line.split()[3][:-3]]
        else:
            key, _, val = line.split()
            data1[key] += [val]
    
def printplot(key1,key2,plotargs=None):
    coords = ''.join(['({0},{1})'.format(x,y)
        for x,y in zip(data1[key1],data1[key2])])
    if plotargs is None:
        plot = r'\addplot coordinates {' + coords + '};'
    else:
        plot = r'\addplot' + plotargs + ' coordinates {' + coords + '};'
    
    print(plot)        

\end{pycode}
\begin{tikzpicture}
\begin{axis}[scaled ticks=false, 
    tick label style={/pgf/number format/fixed}, 
	xtick = {1,400000,800000,1200000,1600000,2000000,2400000,2800000,3200000,3600000},
	xticklabels={,400,800,1200,1600,2000,2400,2800,3200,3600},
	xmin= 0, xmax =3800000,
	ymin=0.52303, ymax=0.52322,
	xlabel=$j$,
	yticklabel style={/pgf/number format/.cd, fixed, /pgf/number format/fixed zerofill,precision=5},
  	ytick={0.52305,0.52310,...,0.52322},
  	ylabel={\tiny Compression ratio},
    cycle list name=exotic,
    mark size = 1.5,
    legend style={draw=none},
    legend style={at={(0.97,0.85)}}
    ]
    \addplot  coordinates {
    (1,0.523213) 
    (50000,0.523187) 
    (200000,0.523162) 
    (400000,0.523123) 
    (600000,0.523098) 
    (800000,0.523088) 
    (1000000,0.523084) 
    (1200000,0.523079) 
    (1400000,0.523075) 
    (1600000,0.523072) 
    (1800000,0.523068) 
    (2000000,0.523063) 
    (2200000,0.52306) 
    (2400000,0.523059) 
    (2600000,0.523058) 
    (2800000,0.523057) 
    (3000000,0.523057) 
    (3200000,0.523057) 
    (3400000,0.523057) 
    (3600000,0.523057) 
    (3800000,0.523057)}; 
\end{axis}
\end{tikzpicture}
\begin{center}
\caption{\sl Compression ratio for $g_j$.}
\label{G_is}
\end{center}
\end{center}
\end{figure}
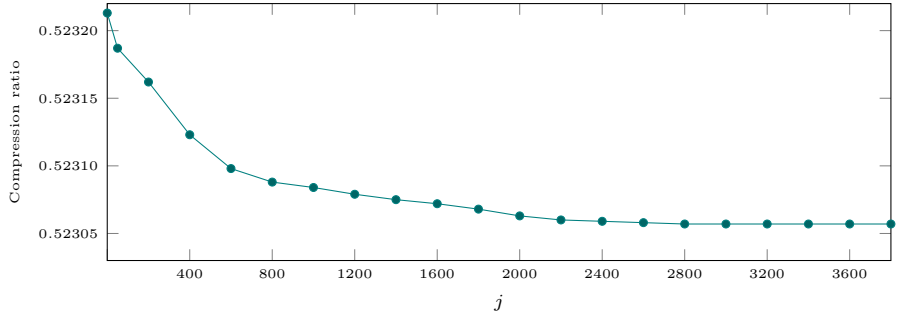

To illustrate the behavior of the family of functions $\cal G$, Figure~\ref{G_is} shows the relative size of the compressed file for selected values of $j$ on our test file to be described below; as expected, the resulting curve is decreasing.

\bigskip

\section{Experimental Results}

To get empirical evidence how the weighted encoding behaves in practice, we considered files of different languages,  sizes and  nature, and obtained similar results. We thus report here only the performance on  the King James version of the English Bible (in which the text has been  stripped of all punctuation signs), as example for typical behavior. 
In order to handle the arithmetic of the huge numbers that are necessary for the coding, we used the GNU Multiple Precision
Arithmetic Library\footnote{\sf https://gmplib.org/}.
Indeed, on some of our tests, the involved numbers used each up to 800 bits.
We applied the weighted compression with two different families of functions, using both Huffman and arithmetic coding.

\begin{figure}[!h]
\begin{center}

\pgfplotsset{
compat=newest,
  width=12cm,
  height=7cm,
}   

\begin{pycode}
from collections import defaultdict
data1 = defaultdict(list)
with open('mydata2.dat') as dd:
    for line in dd:
    	if line.startswith('='):
            if 'k' in line:
                data1['k'] += [line.split()[3][:-3]]
        else:
            key, _, val = line.split()
            data1[key] += [val]
    
def printplot(key1,key2,plotargs=None):
    coords = ''.join(['({0},{1})'.format(x,y)
        for x,y in zip(data1[key1],data1[key2])])
    if plotargs is None:
        plot = r'\addplot coordinates {' + coords + '};'
    else:
        plot = r'\addplot' + plotargs + ' coordinates {' + coords + '};'
    
    print(plot)        

\end{pycode}

\begin{tikzpicture}
[every pin/.style = {pin distance=3mm, 
                font=\scriptsize, inner sep=1pt},
  PN/.style args = {#1/#2}{
    circle, draw=none, fill=red, 
    minimum size=3pt, inner sep=0pt,
    pin=#1:#2}    
]
\begin{axis}[scaled ticks=false, 
    tick label style={/pgf/number format/fixed}, 
	xtick = {0,1,2,4,6,8,10,12,14,16},
	xticklabels={0,1,2,4,6,8,10,12,14,16},
	xmin=0 , xmax =18,
	ymin=0.5222, ymax=0.5234,
	xlabel=$k$,
	ytick={0.5224,0.5226,...,0.5232},
    yticklabel style={/pgf/number format/.cd,fixed, /pgf/number format/fixed zerofill,precision=4},
  	ylabel={\tiny Compression ratio},
    cycle list name=exotic,
    legend style={draw=none},
    legend style={at={(1.0,0.64)}}
    ]
    \addplot  coordinates {
(0,0.523213) 
(0.5,0.523117)
(1,0.523057) 
(1.5,0.523004)
(2,0.522953) (3,0.522873) (4,0.522801) (5,0.522745) (6,0.522693) (7,0.522648) (8,0.522608)
(10,0.52253) (12,0.522459) (14,0.522392) (16,0.52233)}; 
    \addlegendentry{\it \tiny Weighted Coding};
    \addplot  coordinates {
   (0,0.523256) 
   (0.5,0.523181)
   (1,0.523142)
   (1.5,0.523109)
   (2,0.523078) (3,0.523038) (4,0.523005) (5,0.522987) (6,0.522978) (7,0.522971) (8,0.522969)
   (10,0.522969) (12,0.522979) (14,0.52299) (16,0.523005)};
    \addlegendentry{\it \tiny Weighted -- Header+Coding};
\addplot [magenta,dotted,ultra thick, mark=none]
plot coordinates {
            (0,0.523269)
            (16,0.523269)
        };
\addlegendentry{\it \tiny Static -- Header+Coding};
\addplot [gray,dotted,ultra thick, mark=none]
plot coordinates {
            (0,0.523252)
            (16,0.523252)
        };
\addlegendentry{\it \tiny Vitter};

\end{axis}
\end{tikzpicture}
\caption{Compression efficiency of weighted Huffman encoding for $g(i)={(n-i+1)}^k$.}
\label{PosK}
\end{center}
\end{figure}
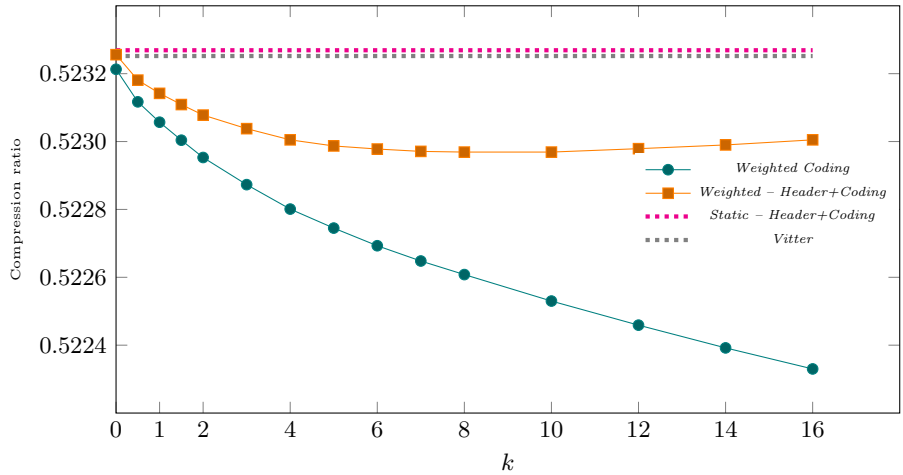

\begin{figure} [!h]
\begin{center}

\pgfplotsset{
  compat=newest,
  width=12cm,
  height=7cm,
}

\begin{pycode}
from collections import defaultdict
data1 = defaultdict(list)
with open('mydata2.dat') as dd:
    for line in dd:
    	if line.startswith('='):
            if 'k' in line:
                data1['k'] += [line.split()[3][:-3]]
        else:
            key, _, val = line.split()
            data1[key] += [val]
    
def printplot(key1,key2,plotargs=None):
    coords = ''.join(['({0},{1})'.format(x,y)
        for x,y in zip(data1[key1],data1[key2])])
    if plotargs is None:
        plot = r'\addplot coordinates {' + coords + '};'
    else:
        plot = r'\addplot' + plotargs + ' coordinates {' + coords + '};'
    
    print(plot)        

\end{pycode}

\begin{tikzpicture}
[every pin/.style = {pin distance=3mm, 
                font=\scriptsize, inner sep=1pt},
  PN/.style args = {#1/#2}{
    circle, draw=none, fill=red, 
    minimum size=3pt, inner sep=0pt,
    pin=#1:#2}    
]
\begin{axis}[scaled ticks=false, 
    tick label style={/pgf/number format/fixed}, 
	xtick = {0,1,2,4,6,8,10,12,14,16},
	xticklabels={0,1,2,4,6,8,10,12,14,16},
	xmin=0 , xmax =18,
	ymin=0.5169, ymax=0.5183,
	xlabel=$k$,
	ytick={0.517,0.5172,...,0.5182},
    yticklabel style={/pgf/number format/.cd,fixed, /pgf/number format/fixed zerofill,precision=4},
  	ylabel={\tiny Compression ratio},
    cycle list name=exotic,
    legend style={draw=none},
    legend style={at={(0.97,0.45)}}
    ]
    \addplot  coordinates {
(0,0.518015) (0.5,0.51791) (1,0.517822) (1.5,0.517747) (2,0.517682) (3,0.517575) (4,0.517491) (5,0.51742) (6,0.517358) (7,0.517302) (8,0.517251) (10,0.517159) (12,0.517076) (14,0.517) (16,0.516929)}; 
    \addlegendentry{\it \tiny Weighted Coding};
    \addplot  coordinates {
   (0,0.518059) (0.5,0.517974) (1,0.517908) (1.5,0.517852) (2,0.517806) (3,0.517741) (4,0.517695) (5,0.517662) (6,0.517643) (7,0.517625) (8,0.517613) (10,0.517598) (12,0.517596) (14,0.517597) (16,0.517604)};
    \addlegendentry{\it \tiny Weighted -- Header+Coding};
\addplot [magenta,dotted,ultra thick, mark=none]
plot coordinates {
            (0,0.518073)
            (16,0.518073)
        };
\addlegendentry{\it \tiny Static -- Header+Coding};
\addplot [gray,dotted,ultra thick, mark=none]
plot coordinates {
            (0,0.518148)
            (16,0.518148)
        };
\addlegendentry{\it \tiny Traditional Adaptive};
\end{axis}
\end{tikzpicture}
\caption{Compression efficiency of weighted arithmetic encoding for $g(i)={(n-i+1)}^k$.}
\label{PosKArith}
\end{center}
\end{figure}
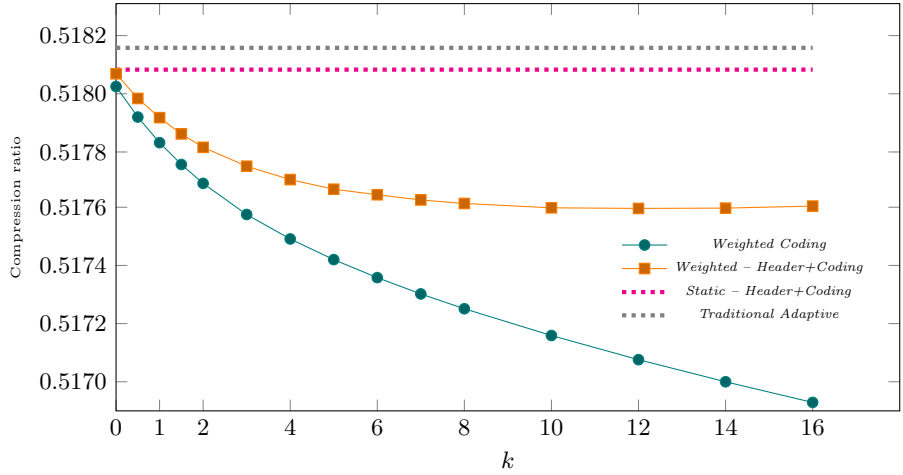

We first considered weighted coding corresponding to functions of the form  $g(i)=(n-i+1)^k$.
Figures \ref{PosK} and~\ref{PosKArith} present the compression ratio, defined as the size of the compressed divided by the size of the original file, for integer values of $k$ ranging from 0 to 16, as well as for $k=0.5$ and $k=1.5$ using both Huffman and arithmetic coding, respectively.
In particular, {\sc Forward} is the special case $k=0$, and {\sc Positional} encoding corresponds to $k=1$.
The lower plot of each graph gives the net encoding while the upper ones include also the necessary header. As can be seen, the compression efficiency improves as $k$ increases in both variants, until about $k=8$ for Huffman, and $k=10$ for arithmetic coding, where the combined (file $+$ header) sizes start to increase, being still better than the size for {\sc Forward}.
The compression ratio for static and dynamic Huffman (Vitter), and static and backward adaptive (Traditional Adaptive) arithmetic coding are given for comparison.

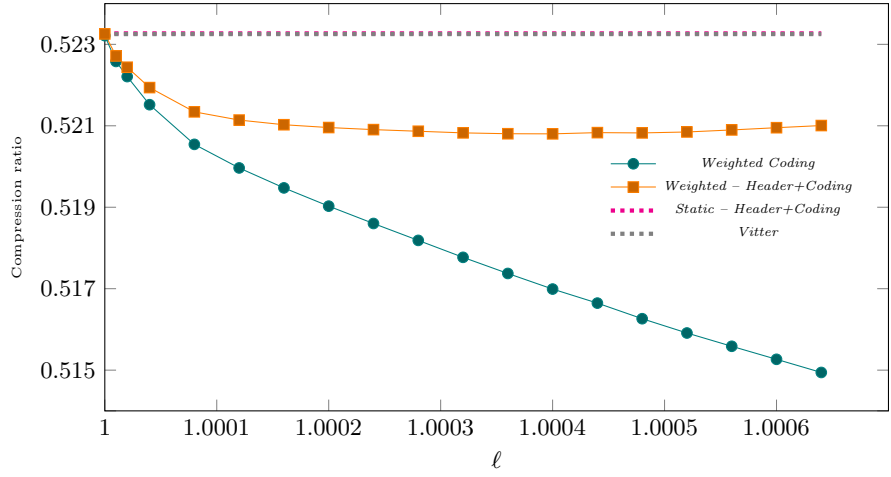
\begin{figure}
\begin{center}

\pgfplotsset{
  compat=newest,
  width=12cm,
  height=7cm,
}

\begin{pycode}
from collections import defaultdict
data1 = defaultdict(list)
with open('mydata2.dat') as dd:
    for line in dd:
    	if line.startswith('='):
            if 'k' in line:
                data1['k'] += [line.split()[3][:-3]]
        else:
            key, _, val = line.split()
            data1[key] += [val]
    
def printplot(key1,key2,plotargs=None):
    coords = ''.join(['({0},{1})'.format(x,y)
        for x,y in zip(data1[key1],data1[key2])])
    if plotargs is None:
        plot = r'\addplot coordinates {' + coords + '};'
    else:
        plot = r'\addplot' + plotargs + ' coordinates {' + coords + '};'
    
    print(plot)        

\end{pycode}
\begin{tikzpicture}
\begin{axis}[scaled ticks=false, 
    tick label style={/pgf/number format/fixed}, 
	xtick = {1,1.0001,1.0002,1.0003,1.0004,1.0005,1.0006},
	xticklabels={1,1.0001,1.0002,1.0003,1.0004,1.0005,1.0006},
	xmin=1.0 , xmax =1.0007,
	ytick={0.514,0.515,0.517,0.519,0.521,0.523},
	yticklabels={,0.515,0.517,0.519,0.521,0.523},
	ymin=0.514, ymax=0.524,
	xlabel=$\ell$,
  	ylabel={\tiny Compression ratio},
    style={at={(2.0,0.5)}},
    cycle list name=exotic,
    legend style={draw=none},
    legend style={at={(0.97,0.65)}}
    ]
    \addplot  coordinates {
(1,0.523213) (1.00001,0.52258) (1.00002,0.522208) (1.00004,0.521518) (1.00008,0.520544) (1.00012,0.519965) (1.00016,0.519475) (1.0002,0.519027) (1.00024,0.518601) (1.00028,0.518186) (1.00032,0.51777) (1.00036,0.517374) (1.0004,0.516991) (1.00044,0.516646) (1.00048,0.516263) (1.00052,0.515912) (1.00056,0.515586) (1.0006,0.515266) (1.00064,0.514943)}; 
    \addlegendentry{\it \tiny Weighted Coding};
    \addplot coordinates {
(1,0.523256) (1.00001,0.522715) (1.00002,0.522438) (1.00004,0.521938) (1.00008,0.521342) (1.00012,0.521141) (1.00016,0.521026) (1.0002,0.520957) (1.00024,0.520906) (1.00028,0.520866) (1.00032,0.520826) (1.00036,0.520805) (1.0004,0.520802) (1.00044,0.520832) (1.00048,0.520825) (1.00052,0.520848) (1.00056,0.520898) (1.0006,0.520953) (1.00064,0.521006)};
    \addlegendentry{\it \tiny Weighted -- Header+Coding};
\addplot [magenta,dotted,ultra thick, mark=none]
plot coordinates {
            (1,0.523269)
            (1.00064,0.523269)
        };
\addlegendentry{\it \tiny Static -- Header+Coding};
\addplot [gray,dotted,ultra thick, mark=none]
plot coordinates {
            (1,0.523252)
            (1.00064,0.523252)
        };
\addlegendentry{\it \tiny Vitter};
\end{axis}
\end{tikzpicture}
\caption{Compression efficiency of weighted Huffman encoding for $g(i)=\ell^{n-i}$.}
\label{KPos}
\end{center}
\end{figure}

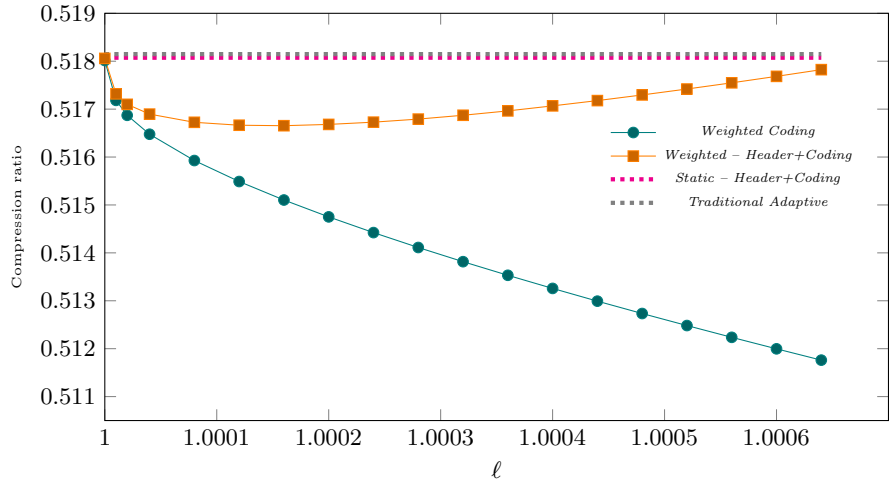
\begin{figure}[!h]
\begin{center}

\pgfplotsset{
  compat=newest,
  width=12cm,
  height=7cm,
}

\begin{pycode}
from collections import defaultdict
data1 = defaultdict(list)
with open('mydata2.dat') as dd:
    for line in dd:
    	if line.startswith('='):
            if 'k' in line:
                data1['k'] += [line.split()[3][:-3]]
        else:
            key, _, val = line.split()
            data1[key] += [val]
    
def printplot(key1,key2,plotargs=None):
    coords = ''.join(['({0},{1})'.format(x,y)
        for x,y in zip(data1[key1],data1[key2])])
    if plotargs is None:
        plot = r'\addplot coordinates {' + coords + '};'
    else:
        plot = r'\addplot' + plotargs + ' coordinates {' + coords + '};'
    
    print(plot)        

\end{pycode}
\begin{tikzpicture}
\begin{axis}[scaled ticks=false, 
    tick label style={/pgf/number format/fixed}, 
	xtick = {1,1.0001,1.0002,1.0003,1.0004,1.0005,1.0006},
	xticklabels={1,1.0001,1.0002,1.0003,1.0004,1.0005,1.0006},
	xmin=1.0 , xmax =1.0007,
	ytick={0.511,0.512,0.513,0.514,0.515,0.516,0.517,0.518,0.519},
	yticklabels={0.511,0.512,0.513,0.514,0.515,0.516,0.517,0.518,0.519},
	ymin=0.5105, ymax=0.519,
	xlabel=$\ell$,
  	ylabel={\tiny Compression ratio},
    style={at={(2.0,0.5)}},
    cycle list name=exotic,
    legend style={draw=none},
    legend style={at={(0.97,0.75)}}
    ]
    \addplot  coordinates {
(1,0.518015) (1.00001,0.517183) (1.00002,0.516871) (1.00004,0.516476) (1.00008,0.515927) (1.00012,0.515489) (1.00016,0.515103) (1.0002,0.514751) (1.00024,0.514423) (1.00028,0.514112) (1.00032,0.513816) (1.00036,0.513532) (1.0004,0.513258) (1.00044,0.512992) (1.00048,0.512734) (1.00052,0.512483) (1.00056,0.512237) (1.0006,0.511997) (1.00064,0.511761)}; 
    \addlegendentry{\it \tiny Weighted Coding};
    \addplot coordinates {
(1,0.518059) (1.00001,0.517319) (1.00002,0.5171) (1.00004,0.516896) (1.00008,0.516725) (1.00012,0.516664) (1.00016,0.516654) (1.0002,0.516681) (1.00024,0.516728) (1.00028,0.516792) (1.00032,0.516872) (1.00036,0.516963) (1.0004,0.517068) (1.00044,0.517178) (1.00048,0.517296) (1.00052,0.51742) (1.00056,0.517549) (1.0006,0.517684) (1.00064,0.517824)};
    \addlegendentry{\it \tiny Weighted -- Header+Coding};
\addplot [magenta,dotted,ultra thick, mark=none]
plot coordinates {
            (1,0.518073)
            (1.00064,0.518073)
        };
\addlegendentry{\it \tiny Static -- Header+Coding};
\addplot [gray,dotted,ultra thick, mark=none]
plot coordinates {
            (1.0,0.518148)
            (1.00064,0.518148)
        };
\addlegendentry{\it \tiny Traditional Adaptive};
\end{axis}
\end{tikzpicture}
\caption{Compression efficiency of weighted arithmetic encoding for $g(i)=\ell^{n-i}$.}
\label{KPosArith}
\end{center}
\end{figure}

The family of functions $g(i)={(n-i+1)}^k$ considered in the first set of experiments does not retain a constant ratio between consecutive positions $i$, which yields a bias towards higher values of $i$.  
For example, referring to our running example of positional Huffman, position 1 is given the weight $p(1)=10$, but the sum of weights from this position on is $\sum_{i=1}^{10}(10-i+1)=55$, so the relative weight for $i=1$ is $10/55=0.18$; on the other hand, $p(8)=3$ yielding a relative weight for position 8 of $3/(3+2+1)=0.5$.
In our following experiment, we thus considered a more balanced family of functions, $g(i)=\ell^{n-i}$, where $\ell$ is a real number slightly larger than 1, which retains a ratio of $\ell$ between consecutive indices.
Figures \ref{KPos} and~\ref{KPosArith} follow the same format as Figures~\ref{PosK} and \ref{PosKArith}.
This time {\sc Forward Huffman} corresponds to $\ell=1$, and again an improvement is achieved.
On this family of functions, the optimal combined size of file plus header is obtained for $\ell=1.0004$ for the Huffman implementation and and for $\ell=1.00016$ for the arithmetic coding implementation.

We see that for both families of weight functions, there is an evident improvement in the compression performance, though only a slight one on the given test file. The significance of our contribution is indeed not the derivation of a ground breaking new compression method, but rather the theoretical and empirical evidence that simple approaches which have been believed to be optimal for years, might at times be improved.

\bibliography{SPIREfull}
\end{document}